\documentclass[conference]{IEEEtran}
\IEEEoverridecommandlockouts
\usepackage{cite}
\usepackage{siunitx}
\usepackage{amsmath,amssymb,amsfonts}
\usepackage{algorithmic}
\usepackage{graphicx}
\usepackage{textcomp}
\usepackage[dvipsnames]{xcolor}
\usepackage{changepage}
\usepackage[inline]{enumitem}
\usepackage{booktabs,multirow}
\usepackage[caption=false]{subfig}

\def\BibTeX{{\rm B\kern-.05em{\sc i\kern-.025em b}\kern-.08em
    T\kern-.1667em\lower.7ex\hbox{E}\kern-.125emX}}
\usepackage[ruled,vlined,linesnumbered]{algorithm2e}
\SetKwInput{KwRequire}{Require}
\usepackage{placeins}
\usepackage{url}

\begin{document}

\title{Array Placement in Distributed Massive MIMO for Power Saving considering Environment Information 
\thanks{This work has been submitted to IEEE International Conference on Communications (ICC2022)}
}

\author{\IEEEauthorblockN{Yi-Hang Zhu, Gilles Callebaut, Liesbet Van der Perre, Fran\c{c}ois Rottenberg}
\IEEEauthorblockA{\textit{
KU Leuven, WaveCore, Department of Electrical Engineering (ESAT)} \\
Ghent Technology Campus, 9000 Ghent, Belgium\\
yihang.zhu@kuleuven.be}
}

\maketitle

\begin{abstract} 

Distributed massive MIMO (D-mMIMO) has been considered for future networks as it holds the potential to offer superior capacity while enabling energy savings in the network.
A D-mMIMO system has multiple arrays.
Optimizing the locations of the arrays is essential for the energy efficiency of the system.
In existing works, array placement has been optimized mostly based on common channel models, which rely on a given statistical distribution and Euclidean distance between user locations and arrays. These models are justified if applied to sufficiently large cells, where the statistical description of the channel is expected to fit its empirical condition. However, with the advent of small cells, this is no longer the case. The channel propagation condition becomes highly environment-specific. 
This paper investigates array placement optimization with
different ways of modeling the propagation conditions taking the environment information (e.g., buildings) into account. We capture the environment information via a graph. Two shortest path-based propagation models are introduced based on the graph. We validate the performance of the models with a ray-tracing simulator considering different signal coverage levels. The simulation results demonstrate that higher energy efficiency is achieved by using the array placements found via the proposed models compared to a Euclidean distance-based propagation model in small cells. The average power saving at 96\% signal coverage, for example, reaches more than \SI{5}{dB}.
\end{abstract}


\section{Introduction}

The concept of the distributed massive multiple-input multiple-output (D-mMIMO) is a marriage of a distributed antenna system~\cite{saleh1987distributed} and massive MIMO~\cite{bjornson2017massive}.
Due to the great potential regarding capacity and energy efficiency, the D-mMIMO has been considered as a key technology for future generation wireless systems~\cite{larsson2014massive}.  
This paper focuses on energy efficient deployment of the D-mMIMO.

A D-mMIMO system relies on the deployment of multiple arrays at different locations to jointly serve the cell users.  
It has been shown that locations for deploying the arrays (array placement) are essential for the energy efficiency of a D-mMIMO system~\cite{zhu2021array}.
Many papers have worked on array placement optimization~\cite{shen2007optimal,wang2009antenna,han2010design,firouzabadi2011optimal,park2012antenna,lin2012optimization,atawia2013indoor,yang2015performance,kamga2016spectral,minasian2018rrh,gopal2020inter,zhang2020optimal,zhu2021array}.  
Most channel propagation models applied in the literature for optimizing the array placement take limited location-specific information into account: large-scale and small-scale fading are modeled based on certain distributions, and the estimated path loss 
is adjusted based on the Euclidean distance between user locations and arrays. 
These models work well for a large coverage region, where the statistical distribution fits the empirical channel condition. 
However, as cell density increases and cell sizes become smaller~\cite{larsson2014massive}, 
empirical channel propagation conditions become very environment-specific, meanwhile, highly depend on the array placement.
Therefore, it can be no longer easily captured by the statistical models.
As suggested by~\cite{lin2012optimization}, {\it ``a propagation model which is based on real-world geographic information is indispensable in the proper optimization of the array placement''}. 
In \cite{lin2012optimization}, they applied a ray-tracing simulator (RTS) for optimizing array placement. 
The RTS accurately captures the environment information. 
However, it requires a long computational time when evaluating the channels. 
Therefore, it is likely infeasible to apply an RTS  
together with an optimization technique such as mathematical programming~\cite{nair2020solving} or meta-heuristic~\cite{talbi2009metaheuristics} for array placement optimization. 

In this paper, we introduce an efficient way of modeling the propagation, while partially capturing the environment information.  
Our goal is to optimize the array placement in a D-mMIMO deployment for reducing the total transmit power.
The contributions of this paper are as follows. 
\begin{itemize}
	\item We formulate the array placement optimization problem for the D-mMIMO system as a mixed integer programming (MIP) formulation. 
	\item We capture the environment information of the considered region via a graph where each node represents a user location in the region. 
	\item 
	We model large-scale fading using a graph-based approach.
	\item Based on an RTS\footnote{Provided by Huawei.}, we validate the performance of the proposed models under different scenarios including different cell sizes and signal coverage levels.
\end{itemize}

In the rest of this paper,
Section~\ref{sec:system} presents the details of the D-mMIMO system that is considered and the MIP formulation we present for the problem. 
Section~\ref{sec:path} elaborates the graph-based approach that we apply to model large-scale fading while capturing the environment information. 
Section~\ref{sec:results} discusses the performance of the different propagation models and Section~\ref{sec:conclusion} concludes the paper. 

In the equations throughout this paper, upper case letters in the default math font are constants. Upper case letters in the mathcal font denote sets. Subscripts in lower case are indices. 

\section{System and Mathematical Formulation}\label{sec:system}
\begin{figure}[tb] 
	\centering
	\includegraphics[width=0.4\textwidth]{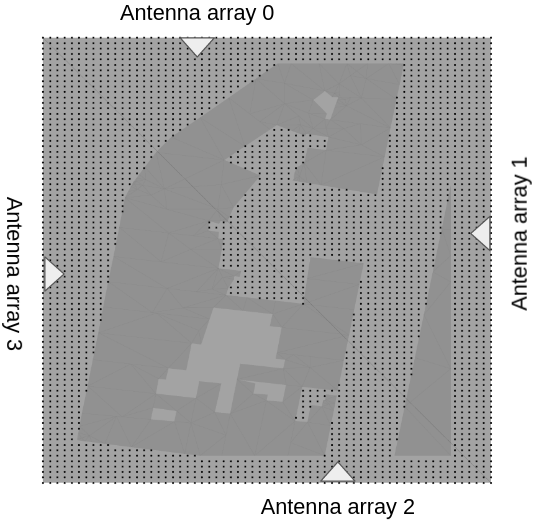}
	\caption{An exemplary D-mMIMO deployment and coverage region. The D-mMIMO system has one array (white triangle) on each edge of the rectangular coverage region. The black dots are the potential user locations that need to be covered by the signal. Light gray indicates the empty space and dark gray indicates the objects in the region, e.g., buildings. The empty space surrounded by the objects is inaccessible, therefore, contains no user location.}
	\label{fig:system}
\end{figure}

Fig.~\ref{fig:system} illustrates an example for the D-mMIMO system and coverage region considered in this paper. 
The considered D-mMIMO system has four arrays, one on each edge of the considered rectangular region (see the white triangles in Fig.~\ref{fig:system}).
Each of the arrays contains $N$ isotropic antennas. 
A single user is assumed to be equipped with a single isotropic antenna.
The black dots in Fig.~\ref{fig:system} illustrate the possible user locations which are generated as follows. 
Initially, the locations are generated such that they are one meter apart and uniformly cover the region. 
Then the locations which are in a building or surrounded by a building are removed.
The user locations on the edges also correspond to the potential locations for deploying the arrays. 

In this work, we reduce the transmit power by optimizing the array locations, while still providing an expected
signal coverage.
We formulate the problem deterministically as 
Eqs.~\eqref{obj:power}-\eqref{constr:coverage}, considering large-scale fading without shadowing effects.


Objective function~\eqref{obj:power} minimizes variable $p_{\mathtt{T}}$, the transmit power per array. 
Constraints~\eqref{constr:one} guarantee that one location is selected for deploying array $t$. 
Parameter $\mathcal{L}_{t}$ in the constraints is the set of all the feasible locations for deploying array $t$.
Parameter $\mathcal{T}$ is the set of all the arrays in the D-mMIMO system.
Variable $x_{tl}$ is binary. 
It equals one if array $t$ is deployed at the location $l$.
We assume that the signals from different arrays are non-coherently combined at each user location.
Constraints~\eqref{constr:power_gain} ensure that the received power at user location $i$ is not lower than the minimum required amount $P_{\mathtt{R}}$.
we assume that through adequate precoding, no user interference is present. 
The minimum received power $P_{\mathtt{R}}$ can be directly translated to a lower bound of the user's channel capacity.
Parameter $\mathcal{R}$ in the constraints is the set of all the user locations in the region. 
Binary variable $z_{i}$ equals one if the received power at user location $i$ is not lower than $P_{\mathtt{R}}$. 
Parameter $\beta_{itl}$ is the channel gain 
received at user location $i$ if array $t$ is deployed at location $l$.

\begin{alignat}{2}
    \label{obj:power}
    &\text{Transmit power minimization:}\nonumber\\
	& \textit{minimize } p_{\mathtt{T}} \\
	&\textit{subject to} \nonumber\\
	&\text{Array placement selection:}\nonumber\\
	\label{constr:one}
	&\sum_{l \in \mathcal{L}_{t}}x_{tl} = 1 
	&\forall t \in \mathcal{T}\\
	&\text{Received power calculation:} \nonumber \\
	\label{constr:power_gain}
	&p_{\mathtt{T}} \sum_{t\in \mathcal{T}}\sum_{l \in \mathcal{L}_{t}} \beta_{itl} x_{tl} \ge P_{\mathtt{R}} z_{i} 
	&\quad\forall i \in \mathcal{R} \\
	&\text{Signal coverage calculation:}\nonumber\\
	\label{constr:coverage}
	&\sum_{i \in \mathcal{R}} z_{i} \ge V_{\mathtt{S}}|\mathcal{R}| 	
\end{alignat}

$\beta_{itl}$ is calculated via Eqs.~\eqref{eq:beta}. 
Parameter $G_{\mathtt{T}}$ in Eqs.~\eqref{eq:beta} denotes the array gain at the transmitter,
equal to the total number of antennas in each of the arrays. 
Parameter $G_{\mathtt{R}}$ is the antenna gain at the receiver, equal to one. 
Parameter $\lambda$ is the wavelength of the considered operating frequency. 
Parameter $d_{itl}$ is the ``{\it distance}'' between user location $i$ and array $t$ if array $t$ is deployed at location $l$. We will incorporate environment information into $d_{itl}$.
More details about $d_{itl}$ are presented in Section~\ref{sec:path}. 

\begin{equation}
	\label{eq:beta}
    \beta_{itl} = \frac{G_{\mathtt{T}}G_{\mathtt{R}}\lambda^{2}}{16\pi^{2}d_{itl}^{2}}
	\quad\forall i \in \mathcal{R}, p\in \mathcal{L}_{t}
\end{equation}

Constraint~\eqref{constr:coverage} ensures the signal coverage, at least $V_{\mathtt{S}}|\mathcal{R}|$ user locations receive sufficient signal power regarding the minimum required amount $P_{\mathtt{R}}$.
$V_{\mathtt{S}}$ is a parameter between zero and one, corresponding to zero signal coverage and full signal coverage, respectively. 
To linearize Eqs.~\eqref{obj:power}-\eqref{constr:coverage}, we divide Constraints~\eqref{constr:power_gain} with $p_{\mathtt{T}}$ and introduce variable $y$, which is related to the transmit power $p_{\mathtt{T}}$ via Eq.~\eqref{eq:transmit_power}. 

\begin{equation}
	\label{eq:transmit_power}
    y = \frac{P_{\mathtt{R}}}{p_{\mathtt{T}}}
\end{equation}
A MIP formulation of the problem is stated as follows.

\begin{alignat}{2}
    \label{obj:power_gain}
	& \textit{maximize } y \\
	&\textit{subject to} \nonumber\\
	&\textit{Constraints~\eqref{constr:one} and \eqref{constr:coverage}} \nonumber\\
	\label{constr:power_gain_mip}
	&\sum_{t\in \mathcal{T}}\sum_{l \in \mathcal{L}_{t}}\beta_{itl} x_{tl} - y \ge M(z_{i}-1) 
	&\quad\forall i \in \mathcal{R} 
\end{alignat}

Objective function~\eqref{obj:power_gain} maximizes variable $y$.
Constraints~\eqref{constr:power_gain_mip} are equivalent to Constraints~\eqref{constr:power_gain}. 
Parameter $M$ is a large value to ensure the validity of the constraints when variable $z_{i}$ equals zero.
After solving the formulation, transmit power $p_{\mathtt{T}}$ can be easily calculated from $y$ regarding Eq.~\eqref{eq:transmit_power}.

\section{Graph based Propagation Models}\label{sec:path}
\begin{figure}[tb]
	\centering
	\includegraphics[width=0.15\textwidth]{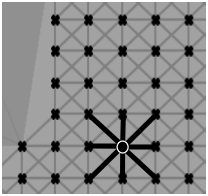}
	\caption{An example of the graph capturing the environment information. This example is based on a part of the cell in Fig.~\ref{fig:system}. The highlighted lines show that each node is connected with maximum eight arcs.}
	\label{fig:graph}
\end{figure}

This section elaborates on how we incorporate the environment information into the parameter $d_{itl}$ in Eqs.~\eqref{eq:beta}. 
We first capture the environment information via a graph. 
Each node in the graph is a potential user location generated in Section~\ref{sec:system} and connected to its neighbor nodes. 
Due to the nature of the graph we applied, each node is associated with at most eight arcs as illustrated by the highlighted example in Fig.~\ref{fig:graph}.
 
Given that we assume each array can only be deployed at a user location that is at the edge of the region,  $d_{itl}$ in Eqs.~\eqref{eq:beta} can be interpreted as the length of the best path between each node and each node on the edges. 
The latter being the potential locations for deploying the arrays. 
The algorithm for computing the best paths is based on Dijkstra’s Algorithm~\cite{dijkstra1959note}.  
In Section~\ref{subsec:path_alg} and~\ref{subsec:angle}, we present the details of the algorithm. 

\subsection{Shortest Path Algorithm}\label{subsec:path_alg}


\begin{algorithm}[!tp]
	\caption{The shortest path algorithm}\label{alg:best_path}
	\KwRequire{$S$, $graph$}
	$dists \leftarrow \emptyset$ \\
	\For{\texttt{each} $source \in S$}{
			$Q \leftarrow \emptyset$ \\
		\texttt{Add} $source$ \texttt{to} $Q$ \\
		\For{\texttt{each node} $n \in graph$}{
			$dist[n] \leftarrow null$ \\
			$prev[n] \leftarrow null$  \\
			$v[n] \leftarrow -\infty$ \\
		}		
		$v[source] \leftarrow 10000$ \\
		$dist[source] \leftarrow 0$

		\While{$Q$ \texttt{not empty}}{
			$i \leftarrow$ $n \in Q$ \texttt{with} $\max v[n]$ \\
			\texttt{Remove} $i$ \texttt{from} $Q$\\
			\For{\texttt{each neighbor} $j$ \texttt{of} $i$ \texttt{in} $graph$}{
				$\gamma = \texttt{factor}(source, i, j, prev)$\\
				$alt = \gamma*v[i] - \texttt{length}(i,j)$ \\
				\If{$alt > v[j]$}{
					\If{$alt < 0$}{
						\texttt{Add} $i$ \texttt{to} $Q$ \\		
						\For{\texttt{each} $n \in graph$}{
							\If{$v[n] \neq -\infty$}{
								$v[n] += 10000$
							}
						}	
						break		
					}
					\texttt{Add} $j$ \texttt{to} $Q$ \\
					$v[j] = alt$\\
					$prev[j] = i$ \\
					$dist[j] = dist[i] + \texttt{length}(i,j)$\\
	 			}
			}
		}
		\texttt{Add} $dist$ \texttt{to} $dists$
	}
	\Return $dists$
\end{algorithm}
We apply Algorithm~\ref{alg:best_path} to search for the paths.  
The intuition behind the search is that when a radiowave travels for a distance, it loses a certain amount of power. 
When a radiowave reaches a location via a mechanism such as diffraction, it also loses power due to the angular change. 
Given an initial power at a source node, the higher the remaining power at a destination node, the better the path for reaching the destination node from the source node.

$graph$ in the algorithm stores all the connections between the nodes. 
Set $S$ includes all the nodes in the graph that are associated with the user locations on the edges of the coverage region, in other words, the potential array locations. 
Lines 2-28 iterate over the nodes in $S$.
For each node  ($source$ node),
the best paths from the $source$ node to each node in the graph (destination nodes) are computed.
In detail, set $Q$ holds the nodes that need to be explored for extending the paths from the $source$ node to other nodes and searching for the best paths from the $source$ node to other nodes (lines 11-27).
Initially, $Q$ only contains a $source$ node (line 4). 
The explored nodes will be removed on line 13 and new nodes will be added to it on line 24.
When $Q$ becomes empty, 
all the best paths are computed for the iteration (line 11). 

$dist[n]$ in the algorithm records the length of the best-found path from the $source$ node to node $n$.
$prev[n]$ records the previous-hop node on the best-found path from the $source$ node to node $n$.
$v[n]$ records the quality of the best-found path from the $source$ node to node $n$, representing the remaining power at node $n$ via the path. 
The greater $v[n]$, the better path for node $n$.
On line 9, a large value, representing the initial power, is assigned to $v[source]$ to ensure that the values in vector $v$ do not easily become negative.

On lines 11-27, line 12 selects node $i$ in $Q$ that has the maximum value in $v$.
Then, all the neighbor nodes of node $i$ are checked. 
Line 15 calculates the factor for the angular changing when moving from node $prev[i]$ to node $i$ and then to node $j$. 
The larger the angular change, the smaller the factor will be.
Section~\ref{subsec:angle} gives the details of the calculation. 
Line 16 calculates the $v$ value for the path from $source$ to node $j$ through node $i$.
The multiplication part represents the power loss due to the angular change.  
The subtraction part represents the power loss due to the distance between nodes $i$ and $j$. 
The calculation on line 16 enables the search for a path with fewer angular changes, particularly at the beginning of the path, and a shorter length. 

If the new value, calculated on line 16, is better than the current $v$ value for node $j$ (line 17), node $j$ will be added to $Q$ on line 24 and the related vectors are updated for $j$ (lines 25-27). 
On line 28, the length of the best-found paths ($dist$) for each $source$ node is stored in $dists$ which is used to construct the parameter $d_{itl}$ in Eqs.~\eqref{eq:beta}. 
Moreover, Lines 18-23 ensure that the values in vector $v$ are non-negative. 
The reason for doing so is that the angular factor $\gamma$ is positive and less than one, the function on line 16 works differently when $v[i]$ is positive and when $v[i]$ is negative.

\subsection{Angular Change Factor}\label{subsec:angle}

\begin{figure}[tb]
	\centering
	\includegraphics[width=0.1\textwidth]{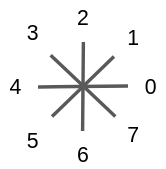}
	\caption{An arc in the graph is associated with one of the eight angles}
	\label{fig:angles}
\end{figure}

\begin{algorithm}[tp]
	\caption{Angular change factor calculation}\label{alg:angle_change}
	\KwRequire{$source$, $i$, $j$, $prev$}
	$\gamma = 1$\\
	\If{$i \neq source$}{
		$\delta = |\texttt{angle}(i,j) - \texttt{angle}(prev[i], i)|$\\	
		$\gamma = \frac{4}{4-\min\{\delta, 8-\delta\}}$ \\
	} 
	\Return $\gamma$
\end{algorithm}
Algorithm~\ref{alg:angle_change} details the procedure for calculating the angular change factor, originating from scattering.  
Line 1 initializes the factor. 
If the path is directly coming from the $source$ node, there will be no angular change and the factor will be one. 
Otherwise, the factor is calculated via lines 3-5.
Function $\texttt{angle}(i,j)$ measures the angle of the arc between nodes $i$ and $j$. 
As illustrated in Figs.~\ref{fig:graph} and~\ref{fig:angles}, there are eight possible angles 0-7 for an arc, corresponding to 0$\times$\ang{45}-7$\times$\ang{45}, respectively.

The reasoning behind lines 3-5 is as follows. 
Considering diffraction for the radiowave
propagation, the maximum angular change is \ang{180} (4$\times$\ang{45}).
Additionally, we assume that the factor equals one when the angular change is \ang{0} and zero when the angular change is 4$\times$\ang{45}(\ang{180}).

\subsection{Three Propagation Models}\label{subsec:models}
We consider three different ways of calculating parameter $d_{itl}$ for Eqs.~\eqref{eq:beta}.
\begin{enumerate}[label=(\alph*)]
	\item Shortest path-based model: parameter $d_{itl}$ equals the length of the best-found path from Algorithm~\ref{alg:best_path} with the angular change factor ($\gamma$) fixed to one.
	\item Shortest path-based model with angular change penalty: parameter $d_{itl}$ equals the length of the best-found path from Algorithm~\ref{alg:best_path} with the angular change factor ($\gamma$) calculated via Algorithm~\ref{alg:angle_change}.
 	\item Euclidean distance-based model: parameter $d_{itl}$ equals the Euclidean distance between the user location and the array. This model represents the common way of modeling the propagation in the literature. We use it for the purpose of comparison. 
\end{enumerate}   


\section{Simulation}\label{sec:results}
In this section, we compare the performance of the three propagation models mentioned in Section~\ref{subsec:models}. 
The parameter setting for the simulations are listed in Table~\ref{tb:para}.
\begin{table}[tb]
\caption{The parameter setting for the experiments}\label{tb:para}
\begin{tabular}{ll}
     \toprule
     Minimum required receiving power ($P_{\mathtt{R}}$) &  \SI{-94}{dBm}\\
     Total number of antennas in each array ($N$) & 16 ($2\times8$) \\
     Different signal coverage levels ($V_{\mathtt{S}}$)& \SI{100}{\percent},...,  \SI{91}{\percent} and \SI{90}{\percent}\\
     Height of each array & \SI{30}{\metre} \\
     Height of each user location & \SI{1.5}{\metre} \\
     \bottomrule
\end{tabular}
\end{table}
Two types of cell sizes are used in the simulation experiments: small and large. 
The edge length for a small cell is around or less than \SI{100}{\metre}, and for a large cell is greater than \SI{200}{\metre}.
The environmental data is exported from \url{www.openstreetmap.org}.

For each signal coverage level and cell, we first calculate the best array placements using the different propagation models. 
Then the placements are evaluated using the RTS. 
During the evaluation, the RTS calculates the average channel gain of a \SI{100}{MHz} OFDM signal via Eq.~\eqref{eq:rts} as $\beta_{itl}$.
In Eq.~\eqref{eq:rts}, $\mathcal{F}$ is the set of sub-carrier frequencies.
As mentioned, each receiver has a single antenna. 
$h_{itmlf}$ is the channel coefficient received at user location $i$ from antenna $m$ of array $t$ when location $l$ is applied for deploying array $t$ and frequency $f$ is used.

\begin{equation}
    \label{eq:rts}
	\beta_{itl} = \frac{\sum_{f\in \mathcal{F}}\sum_{m=0}^{N}|h_{itmlf}|^{2}}{|\mathcal{F}|}
\end{equation}

\subsection{Simplified Cell}
As mentioned in Section~\ref{sec:path}, we use all the user locations generated using the method introduced in Section~\ref{sec:system} for calculating the paths and preparing for the parameter $d_{itl}$.
However, when solving the MIP formulation or running the RTS for validating the results, the large number of user locations will result in a scalability problem regarding memory consumption and long computational time. 


This motivates us to find a way to reduce the total number of user locations for solving the MIP formulation and running the RTS. 
As suggested by our preliminary experimental results, by considering a \textit{simplified cell}, which only has the user locations on the cell edges or next to an object, the total number of user locations is significantly reduced.
What's more, the RTS simulation results remain similar to the case when we keep all the user locations.
Therefore, we consider the simplified cell when solving the MIP formulation and running the RTS for validating the results in the experiments. 

\subsection{Array Placement Optimization Setup}
The MIP formulation introduced in Section~\ref{sec:system} is solved using the commercial solver Gurobi 9.0.3. 
As suggested by our preliminary tests, solving the formulation still requires a long computational time when considering a simplified cell.
Improving the formulation for computational purposes is outside the scope of this paper. 
For reducing the computational time, we reduce the total number of locations in $\mathcal{L}_{t}$ for each array as follows. 
We select possible locations for deploying the arrays on each edge of the rectangular region based on the ratio of the locations to the length of the edge: 0, 0.1,..., 1.
0 and 1 mean the locations at the two corners of the edge.
In other words, $\mathcal{L}_{t}$ now only includes $11$ locations.
Each of the locations is rounded to an integer value, therefore, the locations are also user locations ($\forall t, \mathcal{L}_{t}\subset\mathcal{R}$).
Given that we compare the performance of different propagation models based on the best-found array placements using the models, the comparison remains valid with the aforementioned modification for $\mathcal{L}_{t}$.

\subsection{Small Cell Results}
\begin{figure}[!tb]
	\centering
	\subfloat[Simplified small cell 0: \\the required transmit power]{\includegraphics[width=0.29\textwidth]{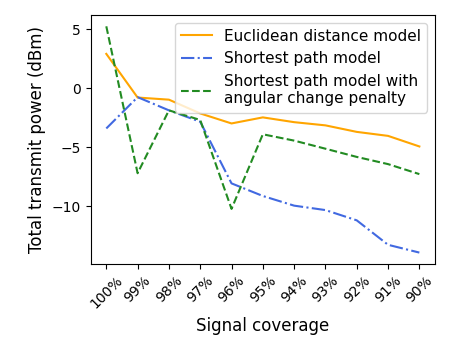}}
	\subfloat[Simplified small cell 0: \\area $\SI{102}{}\times \SI{98}{\metre\squared}$
	]{\includegraphics[width=0.21\textwidth]{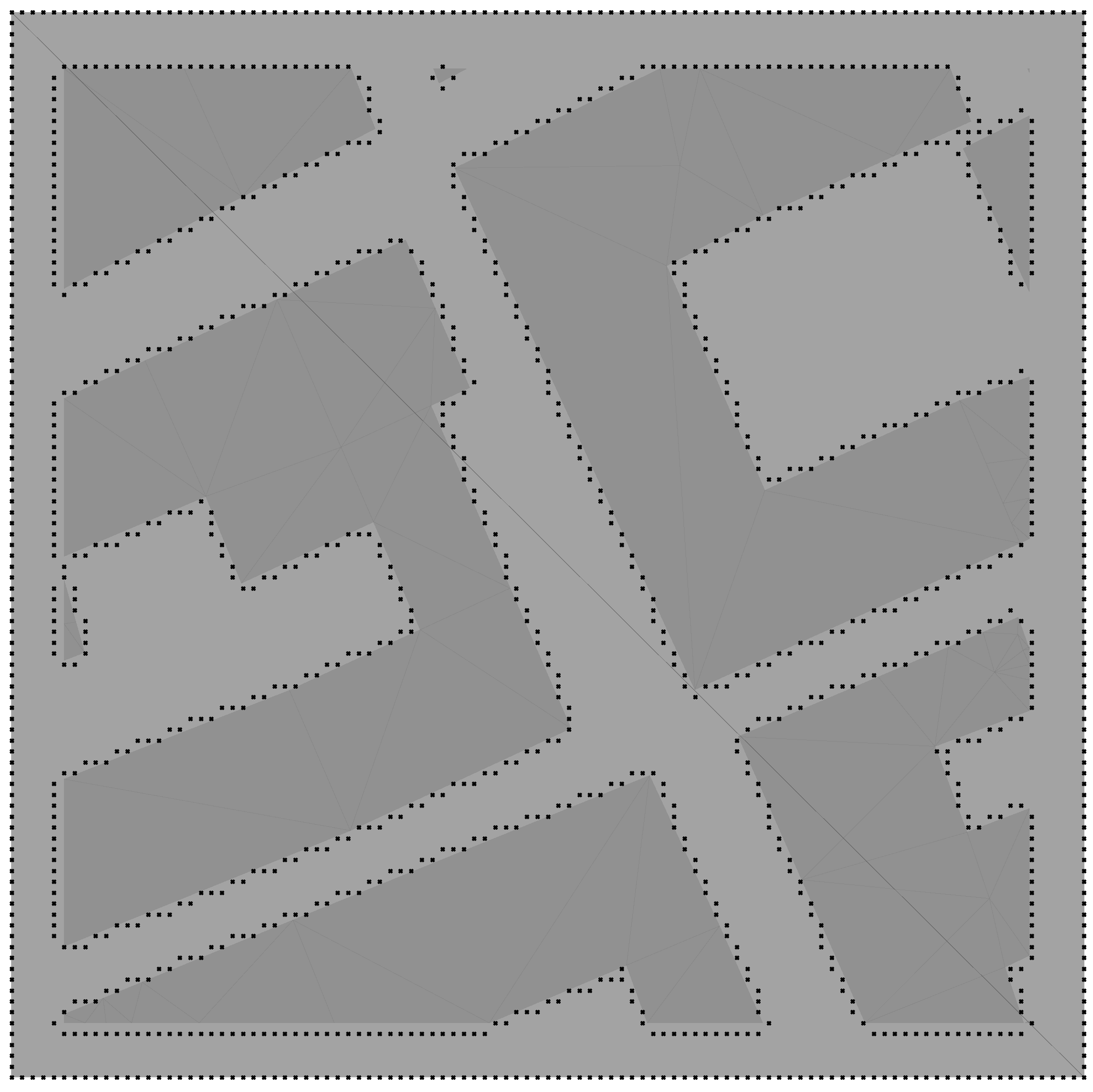}}\\ 
	\subfloat[Simplified small cell 1: \\the required transmit power]{\includegraphics[width=0.29\textwidth]{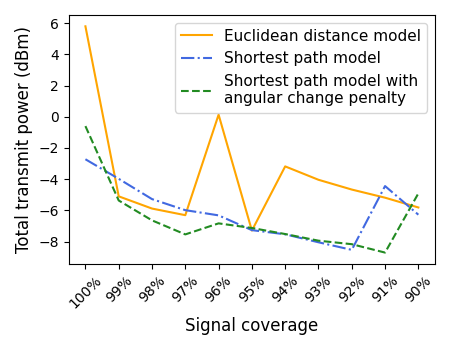}}
	\subfloat[Simplified small cell 1: \\area $\SI{92}{}\times \SI{92}{\metre\squared}$
	]{\includegraphics[width=0.21\textwidth]{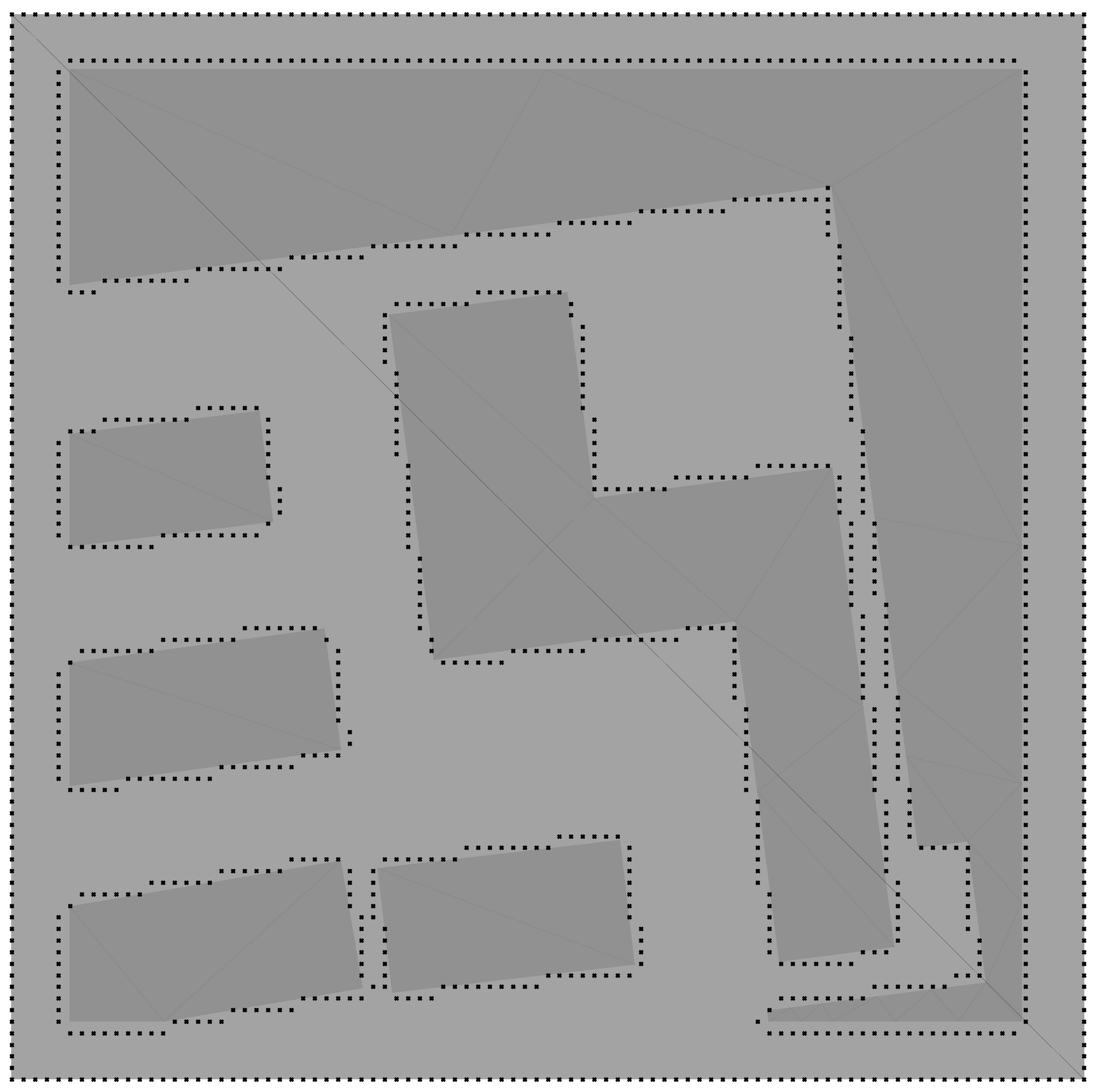}}\\ 
	\subfloat[Simplified small cell 2: \\the required transmit power]{\includegraphics[width=0.29\textwidth]{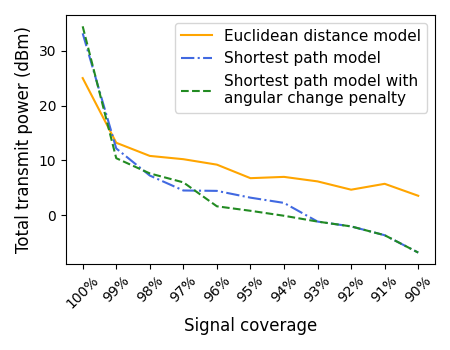}}
	\subfloat[Simplified small cell 2: \\area $\SI{102}{}\times \SI{97}{\metre\squared}$
	]{\includegraphics[width=0.21\textwidth]{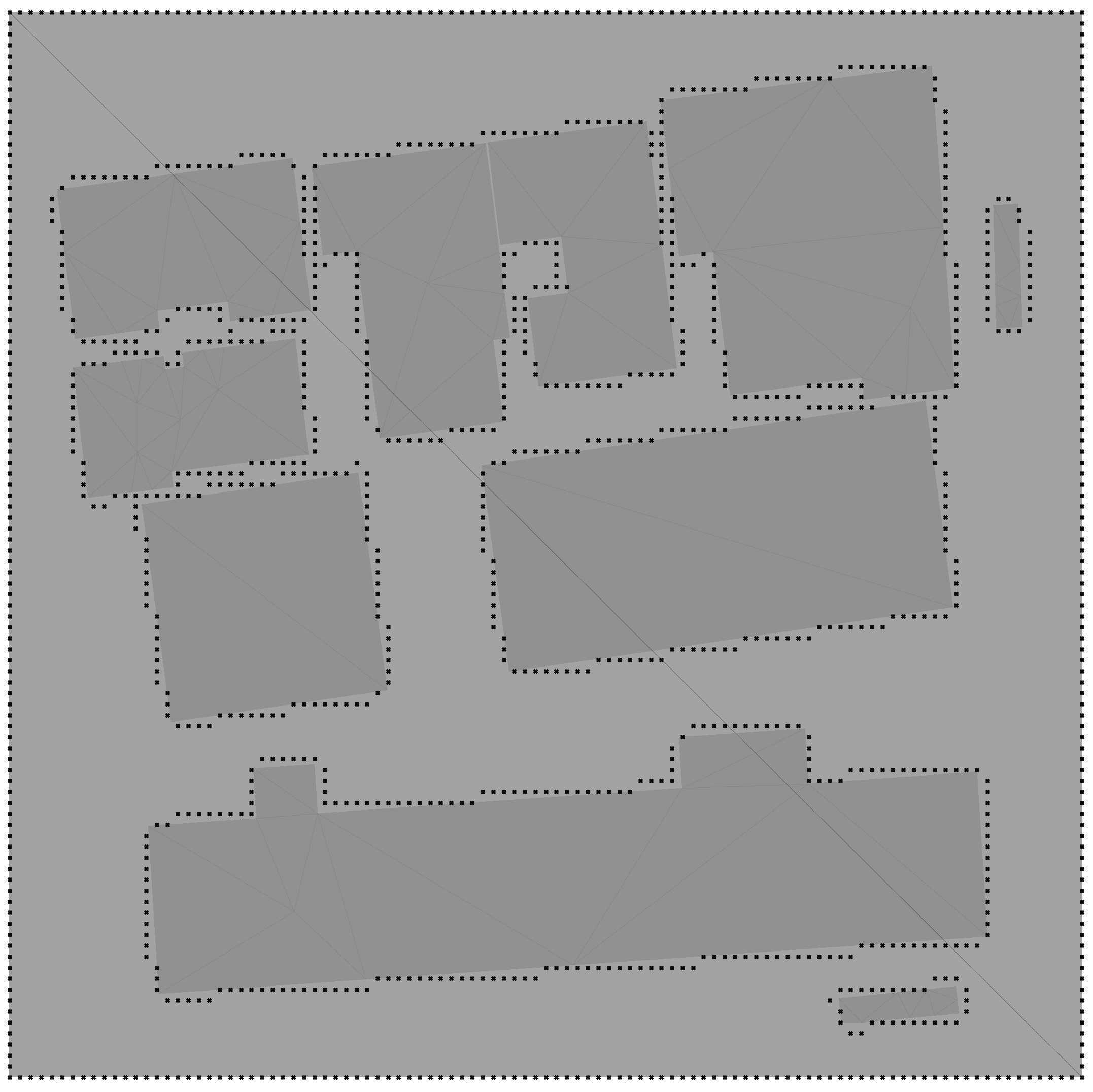}}\\ 
	\subfloat[Simplified small cell 3: \\the required transmit power]{\includegraphics[width=0.29\textwidth]{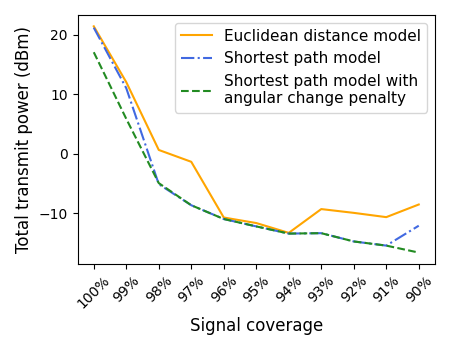}}
	\subfloat[Simplified small cell 3: \\area $\SI{62}{}\times \SI{82}{\metre\squared}$
	]{\includegraphics[width=0.21\textwidth]{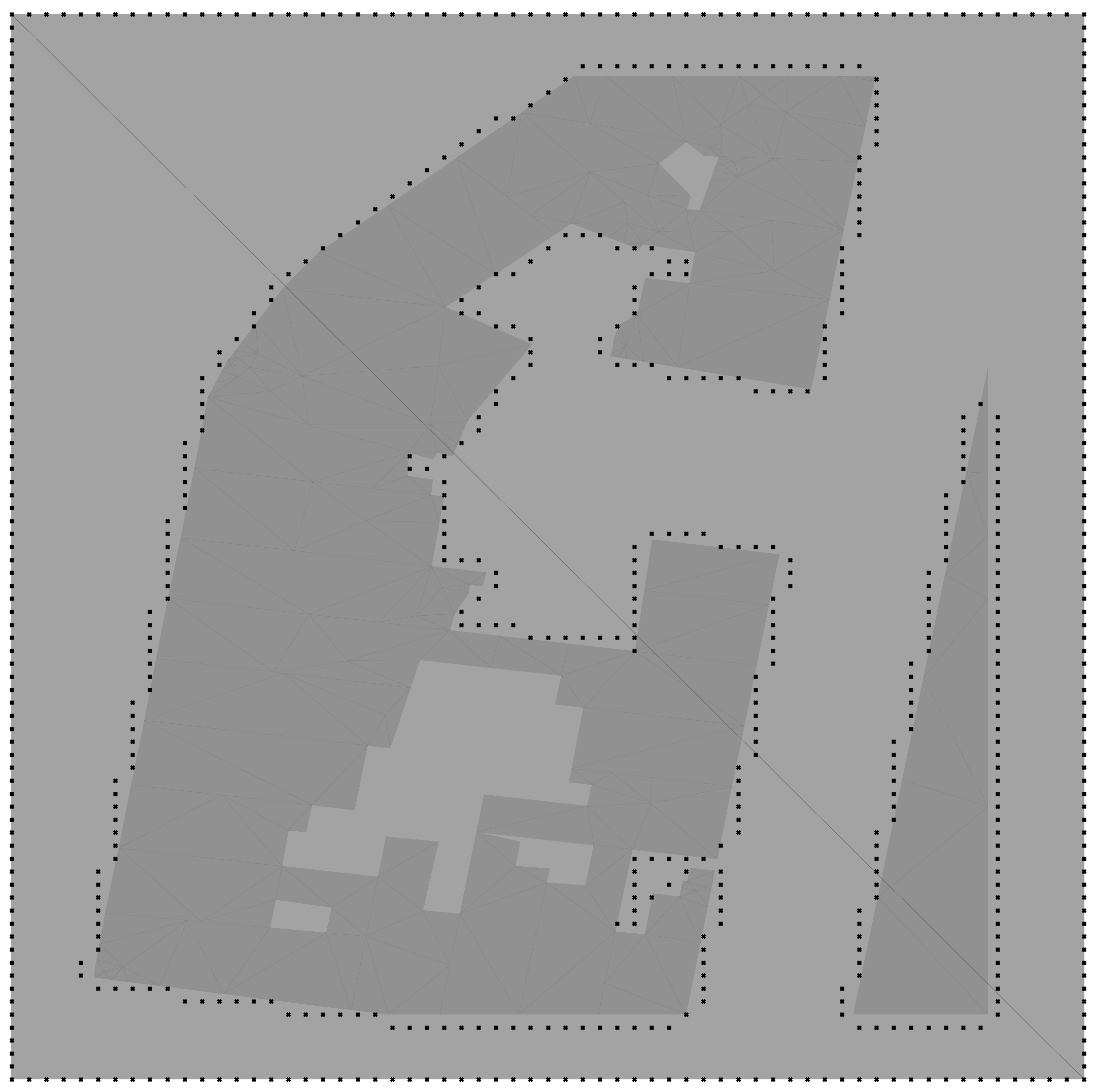}}
	\caption{Comparing the performance of the three propagation models based on the transmit power simulated via the RTS for the best-found placements using the different models in small cells. 
	}
	\label{fig:small}
\end{figure}

The evaluation metric is the transmit power. 
A lower value entails that the model has found better array locations. 
Fig.~\ref{fig:small} illustrates the required transmit power simulated using the RTS for the best placements found by using the three propagation models in small cells. 
The transmit power in Fig.~\ref{fig:small} does not always decrease monotonically as the signal coverage reduces.
The reason is that although the transmit power is calculated based on the array placements which are optimal according to the propagation models, they may turn out to be not optimal choices according to the RTS. 

\begin{figure}
    \centering
    \subfloat[Shortest path-based model]{\includegraphics[width=0.25\textwidth]{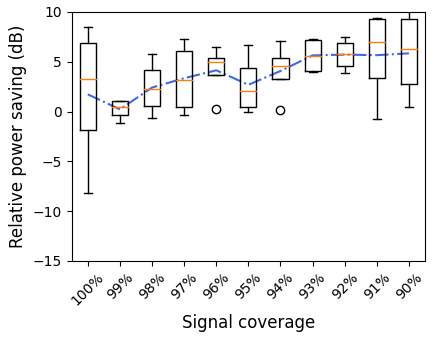}}
    \subfloat[Shortest path-based model with angular change penalty]{\includegraphics[width=0.25\textwidth]{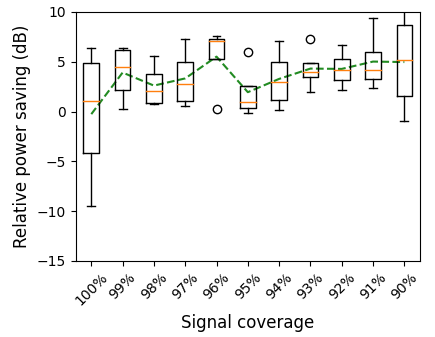}}
    \caption{The relative power saving when using best placements found with the two shortest path-based propagation models compared to the Euclidean distance-based model on the four small cells. Positive values mean that using the best array placement found by a shortest path-based model requires less transmit power. The boxplot is based on the 0th, 25th, 50th, 75th, and 100th percentiles. The dots indicate that the values are outliers. The dash lines connect the mean values of the relative power savings at each signal coverage level on different small cells.}
    \label{fig:relative_saving}
\end{figure}

Fig.~\ref{fig:relative_saving} illustrates the relative power saving when using the best placements found with the two shortest path-based models compared to the best placements found with the Euclidean distance-based model. The positive values in Fig.~\ref{fig:relative_saving} mean that using the placements found by shortest path-based models requires less transmit power. The blue dash lines indicate the mean values for the relative power saving. 

As suggested by Figs.~\ref{fig:small} and~\ref{fig:relative_saving}, though the Euclidean distance-based model sometimes finds better solutions, for most of the cases, both shortest path-based models outperform the Euclidean distance-based model. 
The results confirm that it is important to take the environment information into account when optimizing the array placement in a small cell.
Additionally, Fig.~\ref{fig:small} suggests that, for most of the cases, the shortest path-based model with angular change penalty outperforms the one without angular change penalty. It demonstrates the importance of taking scattering into account for array placement optimization.


\subsection{Large Cell Results}\label{sec:large}
\begin{figure}[tb]
	\centering
	\subfloat[Simplified large cell 0: \\the required transmit power]{\includegraphics[width=0.29\textwidth]{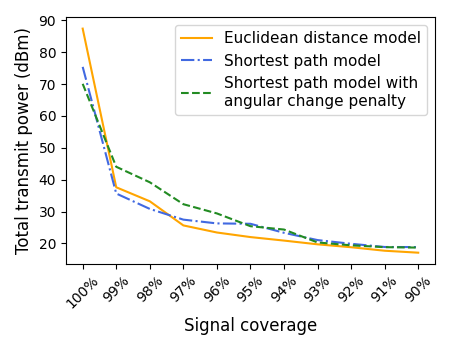}}
	\subfloat[Simplified large cell 0: \\area $\SI{270}{}\times \SI{263}{\metre\squared}$
	]{\includegraphics[width=0.21\textwidth]{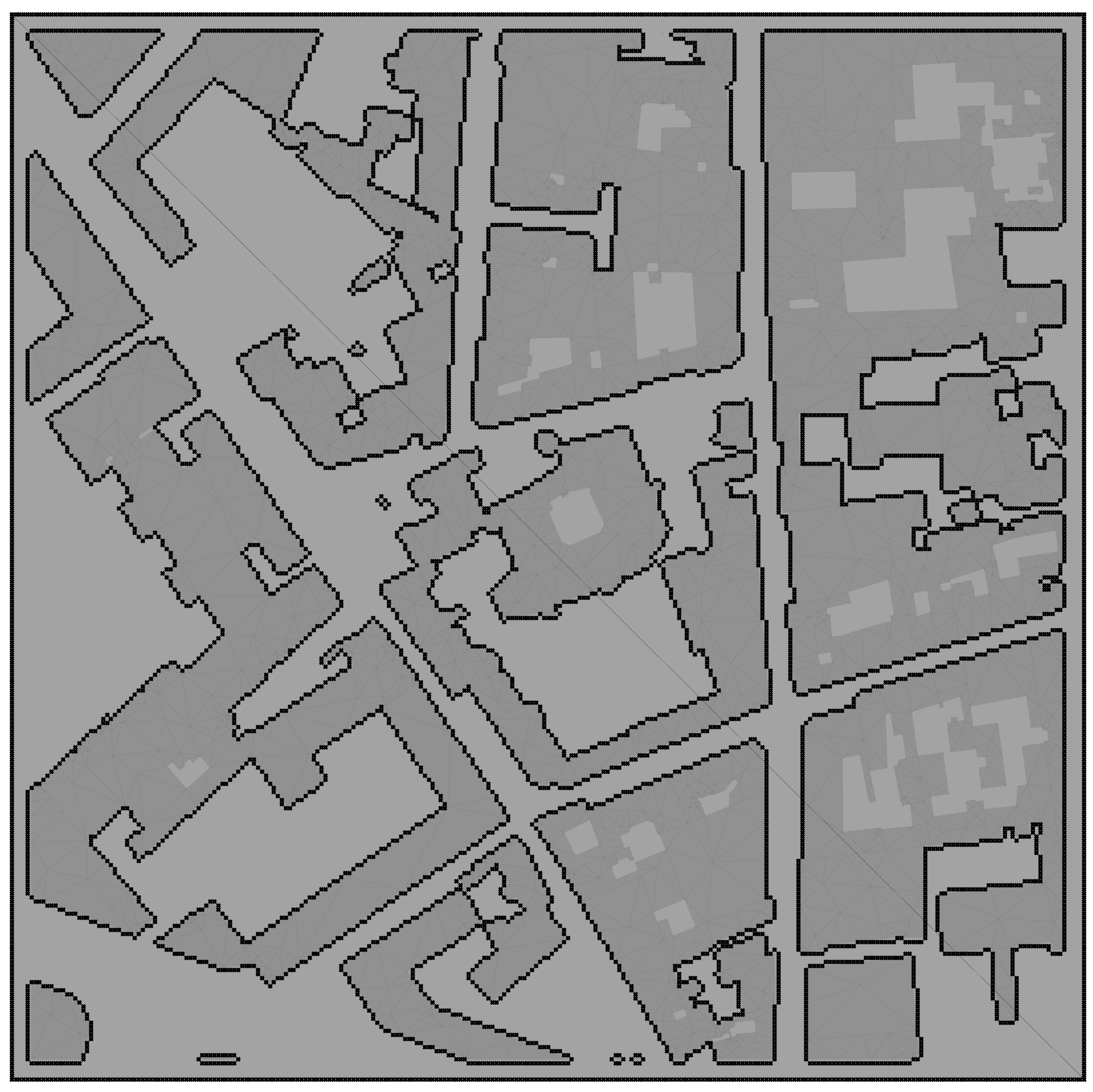}}
	\caption{Comparing the performance of the three different models based on the transmit power simulated via the RTS for the best-found placements using the different models in large cells}
	\label{fig:large}
\end{figure}

Fig.~\ref{fig:large} illustrates the simulated result for the best-found array placement using the different propagation models in a large cell.
As shown in Fig.~\ref{fig:large}, the three models perform similarly for the large cell.
This may be due to that the cell size is large enough such that the location-specific information becomes unimportant.
We will further investigate this perspective in our future work.    


\section{Conclusion}\label{sec:conclusion}
In this paper, we have studied array placement optimization in D-mMIMO deployments to reduce the total transmit power. 
A MIP formulation is presented for the problem.
To better model the channel propagation and generate high-quality array placement for real-world applications of the D-mMIMO, 
we have captured environment information via a graph and introduced two shortest path-based propagation models based on the graph.
The two models are tested considering different cell sizes and signal coverage levels. 
Simulation results demonstrate that it is important to consider environment information when optimizing array placement for the D-mMIMO in small cells. 
Additionally, higher energy efficiency is achieved by using the proposed shortest path-based models for optimizing array placement compared to the Euclidean distance-based model.
The average power saving is more than \SI{5}{dB} at the signal coverage of 96\%.


For future work, validating the propagation models for the case where the signals from different arrays are coherently combined at each user location can be an interesting direction. 
Additionally, this work can also be extended to the case where the arrays have realistic radiation patterns and can be deployed at the locations in the cell (instead of just on the edges). 


\section*{Acknowledgment}
We would like to thank Huawei for their support and NVIDIA for providing the GPU, which has been instrumental in running the RTS. 

\bibliographystyle{IEEEtran}
\bibliography{literature} 

\end{document}